# Flexomagnetoelectric effect in bismuth ferrite

A.K. Zvezdin[1], A.P. Pyatakov[1,2]

1) A.M. Prokhorov General Physics Institute, 38, Vavilova st., Russia, 119991
2) M.V. Lomonosov Moscow State University, Leninskie gory, MSU, Russia, 119992

There is a profound analogy between inhomogeneous magnetoelectric effect in multiferroics and flexoelectric effect in liquid crystals. This similarity gives rise to the *flexomagnetoelectric* polarization induced by spin modulation. The theoretical estimations of flexomagnetoelectric polarization agree with the value of jumps of magnetoelectric dependences (~20μC/m$^2$) observed at spin cycloid suppression at critical magnetic field 200kOe.

**Introduction**

The last few years marked the great progress in the field of magnetoelectric and multiferroics materials [1-3]. The interest to them was triggered by the discovery of so-called spiral multiferroics in which polarization was induced by spin modulation [4-8] and the reports on inverse effects of electrically induced spin modulation [9-12].

These magnetoelectric phenomena can be simply explained by inhomogeneous magnetoelectric interaction proposed in 1980-ies [13] that stems from relativistic exchange of Dzyaloshinskii-Moriya type. However the microscopic mechanism of this coupling is still not clear, and the models proposed recently [14, 15] are being questioned. The supporters of nonrelativistic scenario based on Heisenberg exchange interaction [16] doubt whether weak relativistic coupling is relevant to the electric polarization observed in multiferroics.

In this context the fact of existence of spatially modulated spin structure in bismuth ferrite BiFeO$_3$ gains particular importance. The long range spin cycloid with the period 62nm is known to be induced by spontaneous electric polarization due to the relativistic inhomogeneous magnetoelectric interaction [17]. On the other hand spin modulation may induce an additional electric polarization ΔP due to the same relativistic mechanism [13].

The comprehensive view of magnetoelectric interaction in bismuth ferrite is not only of fundamental but also of practical importance as BiFeO$_3$ is the most promising material for practical application (see reviews [3] and [18] and reference therein). It has record high electric polarization [19,20], and room temperature multiferroic properties. The electric field induced magnetization switching was implemented in BiFeO$_3$/CoFe exchanged coupled structure [21]. There also has been some progress in integrating BiFeO$_3$ with silicon, desirable for Si-CMOS (complementary metal oxide semiconductor) room temperature electronics applications [3].

In this paper the relation between electric polarization and spin modulation in bismuth ferrite is analyzed: to what extent the electric polarization is intrinsic and to what extent it is induced by spin modulation. The anomalies of ferroelectric properties observed in [17] near the magnetic field induced phase transition are reexamined. It is shown that the jump of polarization at phase transition from spin modulated to homogeneous state is the manifestation of additional electric polarization ΔP due to the relativistic mechanism. The profound analogy between inhomogeneous magnetoelectric effect in multiferroics and flexoelectric effect in liquid crystals is emphasized.

**Spatially modulated spin structure**

To find out the role of magnetoelectric interaction in the origins of spin cycloid structure let us examine the contributions to the thermodynamic potential relevant to the magnetic structure. The total expression for the free-energy density has the form

$$F = F_{exch} + F_L + F_{an}, \qquad (1)$$

where

$$F_{exch} = A \sum_{i=x,y,z} (\nabla l_i)^2 = A\left((\nabla \theta)^2 + \sin^2 \theta (\nabla \varphi)^2\right) \qquad (2)$$

is the exchange energy, $l$ is the unit antiferromagnetic (AFM) vector, $A$ the constant of inhomogeneous exchange (or exchange stiffness), and $\theta$ and $\varphi$ are the polar and azimuthal angles of the unit antiferromagnetic vector $\mathbf{l} = (\sin\theta\cos\varphi, \sin\theta\sin\varphi, \cos\theta)$ in the spherical coordinate system with the polar axis aligned with the principal axis $c$, where the second term corresponds to the inhomogeneous magnetoelectric effect in the form of Lifshitz-like invariant:

$$F_L = \gamma \cdot P_s \left(l_x \nabla_x l_z + l_y \nabla_y l_z - l_z \nabla_x l_x - l_z \nabla_y l_y\right). \qquad (3)$$

where $\gamma$ is a constant of inhomogeneous magnetoelectric effect, $P_s$ is spontaneous polarization, $\nabla$ is the differential operator.

and, finally, the third term:

$$F_{an} = -K_u \cos^2 \theta \qquad (4)$$

is the anisotropy energy, and $K_u$ the anisotropy constant.

Minimization of the free-energy functional $F = \int f \cdot dV$ by the Lagrange–Euler method in the approximation ignoring anisotropy gives for the functions $\theta(x,y,z)$ and $\varphi(x,y,z)$

$$\varphi_0 = const = arctg\left(\frac{q_y}{q_x}\right); \quad \theta_0 = q_x x + q_y y; \qquad (5)$$

where q is the wave vector of the cycloid. Equation (5) describes a cycloid whose plane is perpendicular to the basal plane and oriented along the propagation direction of the modulation wavevector.

The exact solution that takes into account anisotropy gives the following expressions for the spin distribution and cycloid period [22,23]:

$$\frac{d\theta}{dx} = \sqrt{\frac{K_u}{A \cdot m}} \sqrt{1 - m\cos^2 \theta} \qquad (6a)$$

$$\lambda = 4K_1(m)\sqrt{\frac{A \cdot m}{K_u}}; \qquad (6b)$$

where $K_1(m) = \int_0^{\pi/2} \frac{d\theta}{\sqrt{1 - m\cos^2 \theta}}$ is an elliptical integral of the first kind, and $m$ the modulus parameter of the elliptical integral that is found by minimization procedure of the free-energy (1) [23]. For an anisotropy constant much smaller than the exchange energy $K_u \ll Aq^2$, the modulus parameter $m$ tends to zero, and solution (6a) becomes harmonic with a linear dependence of $\theta$ on coordinates (5). By substituting (5) into (1), one can obtain the volume-averaged free-energy density for a harmonic cycloid approximation, as

$$\langle F \rangle = Aq^2 - (\gamma P_s)q - \frac{K_u}{2}. \qquad (7)$$

The wave-vector corresponding to the energy minimum is then

$$q_0 = \frac{2\pi}{\lambda} = \frac{\gamma \cdot P_s}{2A}. \qquad (8)$$

Thus inhomogeneous magnetoelectric term (3) gives rise to the spin modulation. The period (6 b) in unperturbed state of cycloid (e.g. in zero external field) can be estimated by formula (8).

## Magnetic field-induced cycloid transformation

Application of high external magnetic fields will result in changes of the effective anisotropy $K_{eff}(H)$ and may disturb or even suppress spin cycloid.

### $H \| c$ axis geometry

Consider the case of $\mathbf{H} \| c$-axis. It is convenient to use dimensionless units of magnetic field:

$$h = H\sqrt{\frac{\chi_\perp}{2Aq_0^2}};  \quad (9)$$

where $\chi_\perp$ the magnetic susceptibility in the direction perpendicular to the antiferromagnetic vector $\mathbf{l}$, $q_0$ (8) the value of the wavevector of the cycloid corresponding to the minimum of the free energy (1) in the absence of external fields and neglecting anisotropy.

The free energy density can be conveniently written as the sum

$$f = f_{exch} + f_L + f_{an}, \quad (10)$$

where the energies of exchange, inhomogeneous magnetoelectric interaction, and effective anisotropy are normalized to the exchange energy $Aq_0^2$ of the harmonic cycloid in the absence of applied fields. That is,

$$f_{exch} = \frac{F_{exch}}{Aq_0^2} = \frac{1}{q_0^2}\left(\frac{d\theta}{dx}\right)^2 \quad (11)$$

$$f_L = \frac{F_L}{Aq_0^2} = -\frac{2}{q_0} \cdot \left|\frac{d\theta}{dx}\right| \quad (12)$$

$$f_{an} = -k(h)\cos^2\theta; \quad (13)$$

where $k(h) = \frac{K_{eff}}{Aq_0^2} = (k_u - h^2 - \beta)$ is the dimensionless effective anisotropy constant, which takes into account the effect of magnetic field $h$, and weak ferromagnetism induced by spontaneous electric polarization: $\mathbf{M}_0 = \alpha[\mathbf{P}_s \times \mathbf{l}]$ that is expressed by the parameter $\beta = \frac{M_0^2}{2\chi_\perp Aq_0^2}$. For the homogeneous state with an antiferromagnetic vector $\mathbf{l} \perp c$ ($\theta = \frac{\pi}{2}$) we have

$$f_\perp = 0. \quad (14)$$

In the spin cycloid state, we can obtain the expression for the total free energy (10) averaged over the period $\langle f \rangle_\lambda = \frac{4}{\lambda}\int_0^{\pi/2} f(\theta)\frac{dx}{d\theta}d\theta$ by taking into account (6a):

$$\langle f \rangle_\lambda = -\frac{k(h)}{m}\left(1 - 2\frac{K_2(m)}{K_1(m)}\right) - \frac{\pi}{K_1}\sqrt{\frac{k(h)}{m}}; \quad (15)$$

where $K_1(m) = \int_0^{\pi/2} \frac{d\theta}{\sqrt{1-m\cos^2\theta}}$, $K_2(m) = \int_0^{\pi/2} \sqrt{1-m\cos^2\theta} \cdot d\theta$ are elliptical integrals of the first and second kind, respectively.

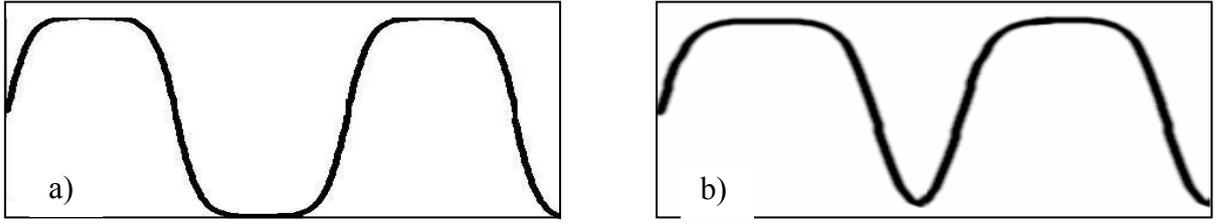

Fig 1. The transformation of the cycloid profile $L_x(x)$ in the high magnetic field: a) h||c b) h ⊥c.

To each field strength $h$ there corresponds a modulus $m$ of the elliptic integral for which the energy is minimum. Physically, this means that the cycloid's profile changes under applied field (fig 1 a). Under strong field, its shape differs significantly from that of a harmonic profile, becoming similar to a function describing a periodic structure of domains separated by walls (solitons) whose widths are considerably smaller than the domain width. It follows from (14) that, upon the transformation to the phase with the antiferromagnetic vector $l \perp c$, the energy of the domain walls $\langle f \rangle_\lambda$ changes sign and the spatially modulated spin state becomes energetically unfavorable.

*H⊥ c axis geometry*

In magnetic field $H \perp c$ in free energy (10) the additional term appears that corresponds to Zeeman energy in magnetic field:

$$f_{Zeeman} = -\frac{M_0 H \sin\theta}{A q_0^2} = -2\sqrt{\beta} h \sin\theta \tag{16}$$

The expressions (6 a) for the spin distribution in cycloid is also modified:

$$\frac{d\theta}{dx} = \sqrt{\frac{K_u}{A \cdot m}} \sqrt{1 - m\cos^2\theta - \tilde{m}\sin\theta}, \tag{17}$$

where $\tilde{m} = m\frac{\sqrt{\beta} \cdot h}{k(h)}$ is the parameter that characterizes asymmetry of the spin cycloid: the in-plane directions of antiferromagnetic vector are not equivalent any more. The direction of $l$ that corresponds magnetoelectrically induced magnetization $\mathbf{M}_0 = \alpha[\mathbf{P}_s \times \mathbf{l}]$ oriented parallel to the external magnetic field is energetically more preferable than the one with magnetization oriented antiparallel to the field.

Formulas for the period (6b), and for the exchange (11), inhomogeneous magnetoelectric interaction (12) and anisotropy (13) contributions as well as averaged total energy remain valid provided that the elliptic integrals are replaced with the following ones:

$$K_1 = \frac{1}{4}\int_0^{2\pi} \frac{d\theta}{\sqrt{1 - m\cos^2\theta - 2\sqrt{\beta}\frac{m}{k(h)}h_x \sin\theta}} \tag{18a}$$

$$K_2 = \frac{1}{4}\int_0^{2\pi} \sqrt{1 - m\cos^2\theta - 2\sqrt{\beta}\frac{m}{k(h)}h_x \sin\theta} \cdot d\theta, \tag{18b}$$

and the energy of homogeneous state (14):

$$f_\perp = -2\sqrt{\beta} h \tag{19}$$

In high field h⊥c the shape of the cycloid differs significantly from that of a harmonic profile. Unlike the case of h||c (fig 1a) the domains are not equal: the ones parallel to the external magnetic field shrink while those ones that have the magnetization antiparallel to the external field inflate (fig 1 b).

**Flexomagnetoelectric effect**

There is a profound analogy between spatially modulated structures in a ferroelectromagnet and waves of the director vector a nematic liquid crystal [13,24,25]. This correlation formally manifests itself in the similarity of the expression for the inhomogeneous magnetoelectric interaction (3) in a multiferroic and the one for the flexoelectric effect in liqiud crystals:

$$F_{Flexo-electric} = \gamma \cdot \mathbf{E}\left[\mathbf{n}(\nabla \mathbf{n}) - (\mathbf{n} \cdot \nabla)\mathbf{n}\right] \quad (20)$$

It can be easily shown that (20) is isotropic form of (3) provided that the director vector **n** stands for antiferromagnetic one ***l*** and external electric field **E** stands for spontaneous electric polarization $P_s$. This is gives us the grounds to name the energy term (3) as *flexomagnetoelectric interaction* and additional polarization ΔP induced by spin modulation as *flexomagnetoelectric* one:

$$\Delta P = -\frac{\partial F_L}{\partial E} = \gamma \kappa \frac{d\theta}{dx} \quad (21)$$

where $\kappa$ is electric susceptibility of the material: $P = \kappa E$, $E$ is electric field.

For the averaged over the period flexomagnetoelectric polarization we obtain simple expression:

$$\langle \Delta P \rangle = \frac{1}{\lambda} \int_0^{2\pi} \Delta P(x) \frac{dx}{d\theta} d\theta = \frac{2\pi}{\lambda} \gamma \kappa \quad (22)$$

where λ is a period of the cycloid determined by (6b).

In figure 2 the numerically calculated wavelength (6b) and electric polarization (22) in dimensionless units are given.

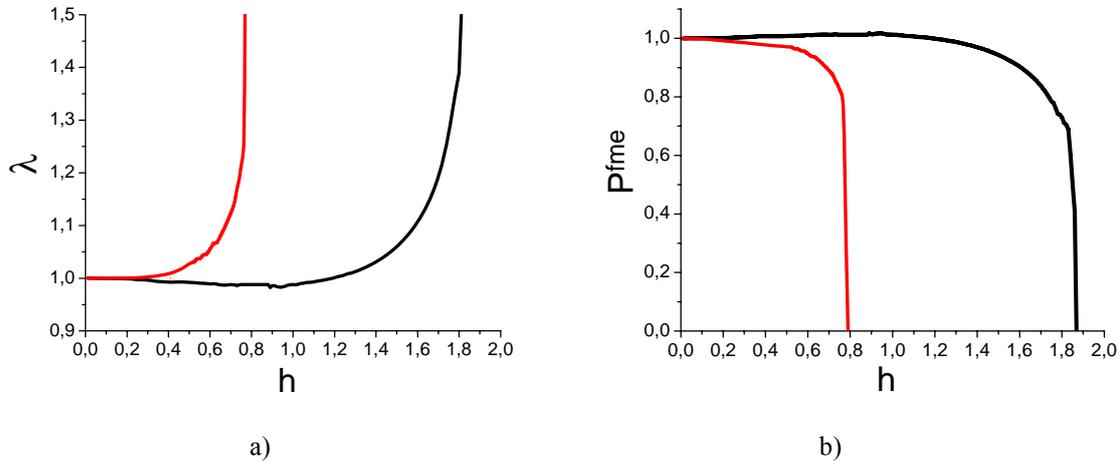

a)          b)

Fig 2 a) the magnetic field dependence of the cycloid period normalized on its zero field value $\lambda_0$=62nm b) the dependence of normalized electric polarization near phase transition to homogeneous state calculated from the magnetic field dependence of cycloid period. Black line is for h∥c axis, red is for h⊥c axis

From (8) and (22) we can estimate the flexomagnetoelectric polarization that should result in anomaly in magnetic field dependence of electric polarization:

$$\langle \Delta P \rangle = \frac{2\kappa A q^2}{P_s}; \quad (22)$$

Taking into account $P_s$~1C/m$^2$ (3 10$^5$ CGS) [19,20], $A = 3 \cdot 10^{-7}$ erg/cm, $q_0 = 10^6$ cm$^{-1}$, $\kappa = \frac{\varepsilon}{4\pi} - 1 \approx 3$ we obtain for ΔP~2 10$^{-5}$ C/m$^2$ (6 CGS) that is very close to the value of polarization jump observed in magnetoelectric dependence $\Delta P_c(H_c)$ near the field h=1.9 (H~200kOe) of phase transition from spin modulated to homogeneous state [17].